\documentclass[12pt,preprint]{aastex}

\usepackage{natbib}

\def\VS{V\'azquez-Semadeni}

\shorttitle{Density PDF}
\shortauthors{Gazol \& Kim}

\begin{document}


\title{The Density Distribution in Turbulent \\
        Bi-stable Flows}


\author{Adriana Gazol\altaffilmark{1}}
\affil{Centro de Radioastronom{\'\i}a y Astrof{\'\i}sica, UNAM, A.
P. 3-72, c.p. 58089, Morelia, Michoac\'an, M\'exico }
\email{a.gazol@crya.unam.mx}

\and

\author{Jongsoo Kim\altaffilmark{2}}
\affil{Korea Astronomy and Space Science Institute, 61-1,
Hwaam-Dong, Yuseong-Ku Daejeon 305-348, Korea}
\email{jskim@kasi.re.kr}

\begin{abstract}
We numerically study the volume density probability distribution function ($n$--PDF) and the column density probability distribution function ($\Sigma$--PDF) resulting from thermally bistable turbulent flows. We analyze three-dimensional hydrodynamic models in periodic boxes of 100$\,$pc by side, where turbulence is driven in the Fourier space at a wavenumber corresponding to 50$\,$pc. 
At low densities ($n\lesssim 0.6\,$cm$^{-3}$) the $n$--PDF, is well described by a lognormal distribution for average local Mach number  ranging from $\sim$0.2 to $\sim$5.5. As a consequence of the non linear development of thermal instability (TI), the logarithmic variance of the distribution for the diffuse gas increases with $M$ faster than in the well known isothermal case. The average local Mach number for the dense gas ($n\gtrsim 7.1\,$cm$^{-3}$) goes from $\sim$1.1 to $\sim$16.9 and the shape of the high density zone of the $n$--PDF changes from a power-law at low Mach numbers to a lognormal at high $M$ values. In the latter case the width of the distribution is smaller than in the isothermal case and grows slower with $M$. At high column densities the $\Sigma$--PDF is well described by a lognormal for all the Mach numbers we consider and, due to the presence of TI, the width of the distribution is systematically larger than in the isothermal case but follows a qualitatively similar behavior as $M$ increases. Although a relationship between the width of the distribution and $M$ can be found for each one of the cases mentioned above, these relations are different form those of the isothermal case.
\end{abstract}



\section{Introduction}\label{sec:intro}
The density distribution and particularly the density Probability Distribution
Function (PDF) has become a crucial ingredient in theories about molecular
cloud and core formation and evolution and on star formation theories (e.g.
Padoan \& Nordlund 2002; Krumholz \& McKee 2005; Elmegreen 2002, 2008; Hennebelle \& Chabrier 2008; Elmegreen 2011; Padoan \& Nordlund 2011; Zamora et al. 2012).

The volume density PDF in turbulent compressible flows has been widely studied for the isothermal case. In particular, numerical experiments of driven and decaying turbulence (V\'azquez-Semadeni 1994; Padoan et al. 1997; Stone et al. 1998; Klessen 2000; Ostriker et al.2001; Boldryev et al. 2002; Beetz et al. 2008; Federrrath et al. 2008) have shown that this distribution has a lognormal shape in a large number of situations. 
From the theoretical point of view, the development of a lognormal distribution for isothermal flows have been explained as a consequence of the multiplicative central limit theorem assuming that individual density perturbations are independent and random (V\'azquez-Semadeni 1994; Passot \& V\'azquez-Semadeni  1998; Nordlund \& Padoan 1999). Furthermore, Passot \& V\'azquez-Semadeni (1998) showed that the density PDF develops a power-law tail at high (low) densities for values of the polytropic index $\gamma$, smaller (larger) than 1.
Also for the isothermal case, numerical simulations at a fixed rms Mach number $M$, show an empirical relationship between the width of the density PDF, as measured by its variance or its standard deviation $\sigma$ and $M$ (e.g. Padoan et al. 1997,  Passot \& V´azquez-Semadeni 1998). For the distribution of $\ln n/\bar n$ this relationship reads 
\begin{equation}
\sigma_s^2=\ln (1+b^2M^2),
\label{sigmamach3d}
\end{equation}
where $b$ is a constant of the order of unity whose value is not clearly established. In the literature it goes from $0.26$ (Kritsuk et al. 2007) to 1 (Passot \& V\'azquez-Semadeni 1998) and seems to vary depending on the relative degree of compressible and solenoidal modes of the forcing (Federrath et al. 2008, Federrath et al. 2010). Recently Price et al. (2011) found that for $M$ up to 20 $b=1/3$ is a good fit to numerical simulations with solenoidal forcing and is in agreement with other recent numerical results (Federrath et al. 2008, 2010; Lemaster \& Stone 2008).

On the other hand, the condensation of diffuse gas produced by the isobaric mode of thermal instability (TI; Field 1965) triggered in colliding flows has been recognized as playing an important role for molecular cloud formation (e.g. Hennebelle \& P\'erault 1999; V\'azquez-Semadeni et al. 2006; Heitsch \& Hartman 2007), specially in early phases. However  the density  PDF resulting in turbulent bi-stable flows, which preserves the bi-modal nature that is signature of the development of TI, and the behavior of the distribution for each phase  has not been systematically studied. 

The density PDF resulting from non-isothermal simulations, has been nevertheless reported in some papers.   Using two-dimensional simulations of turbulent thermally bistable flows Gazol et al.(2005) have shown that the effective polytropic index of the gas increases with  $M$, and that for Mach numbers between 0.5 and 1.25 (with respect to the gas at $10^4$~K) it remains $<1$, with specific values depending also on the forcing scale going from $\approx 0.2$ to $\approx 0.4$ for large scale forcing. They also found that the bimodal nature of the distribution becomes less pronounced as the $M$ increases. The same behavior has been reported from three dimensional simulations (Seifried et al. 2011). Audit \& Hennebelle (2010) compare the density distribution resulting from a bi-stable simulation and the one resulting when a single polytropic equation of state with $\gamma=0.7$ is used. In the latter case they find that the low density part of the PDF is well described by a lognormal distribution while for the higher densities the PDF is a power-law with an exponent of about -1.5, which is consistent with the distribution expected for a very compressible gas with $\gamma<1$ predicted by Passot \& V\'azquez-Semadeni (1998). For the cooling run they find that the density distribution of the cold gas, whose effective polytropic index is close to $0.7$, is well fitted by a log-normal for values larger than $\sim 300\,$cm$^{-3}$. They attribute the difference in the behavior to possible resolution effects and rise the question about whether or not the choice of log-normal distribution in models of molecular clouds is adequate.

The density distribution resulting from more complex systems including a variety of physical ingredients has been studied by a large number of authors. Here we just mention some examples that illustrate the activity in this direction.  
In the context of galactic discs, the effect of varying the polytropic index has been numerically studied by  Li et al. (2003) who find that for self-gravitating gas the density distribution shows an imperfect lognormal shape whose width decreases as $\gamma$ increases. 
Wada \& Norman (2007) use three-dimensional hydrodynamic simulations of a globally stable, inhomogeneous cooling ISM in galactic disks and report density PDFs which are well fitted by a single lognormal function (with larger dispersions for more gas-rich systems) over a wide density range. However,  Robertson \&  Kravtsov (2008), include a detailed model of cooling and heating for temperatures $<10^4\,$K and account for the equilibrium abundance of H$_2$. They conclude that each thermal phase in their model galaxies has its own lognormal density distribution, implying that using a single lognormal pdf to build a model of global star formation in galaxies is likely an oversimplification. Density distributions with well defined peaks have also been obtained from models of vertically stratified interstellar medium dominated by supernovae (e.g. De Avilez \& Breitschwerdt 2005; Joung \& MacLow 2006). 

Observationally, PDFs of the average volume density in diffuse interstellar gas ($n_{\rm HI}<1\,$cm$^{-3}$ for the atomic gas) have been reported by  (Berkhuijsen \& Fletcher 2008). For the atomic gas they used 375 lines of sight from the sample given by Diplas \& Savage (1994) and they estimate the volume density using the column density and the distance to each star. They find that the PDF tends to be a log-normal but cannot be well fitted by a single curve. In particular different widths and means are obtained depending on the position (in the disk or away from it) and on the type of sampled gas. Specifically, they find that a mixture of cool and warm gas along the LOSs causes an increase in the dispersion. 

 An associate tool is the column density PDF ($\Sigma$--PDF), which has been numerically studied mainly in the context of isothermal turbulent flows. V\'azquez-Semadeni \& Garc{\'\i}a (2001) studied the relationship between a lognormal volume density PDF ($n$--PDF) and the resulting column density PDF, finding that when the number of decorrelated density structures along the line of sight is small enough, then the $\Sigma$--PDF is representative of the $n$--PDF and do also have a lognormal shape. As the number of decorrelated density structures increases the column density PDF slowly transits to a Gaussian distribution passing through an intermediate stage where the  distribution shows an exponential decay. As the relationship between the $n$--PDF and the $\Sigma$--PDF depends on the applicability of the Central Limit Theorem, the authors argue that for non-isothermal flows having a volume density PDF with a well defined variance, this theorem should apply and the $\Sigma$--PDF should converge to a Gaussian for lines of sight with a large enough number of decorrelated density structures. More recently, Burkhart \& Lazarian (2012) used solenoidaly driven isothermal MHD simulations to investigate the presence of an empirical relationship between the variance of the column density distribution and the sonic Mach number. They found a relationship with the same form as equation (\ref{eqn:den}), namely
\begin{equation}
\sigma_{ln (\Sigma/\Sigma_0)}^2=A_{\Sigma}\ln (1+b_{\Sigma}^2M^2),
\label{eqn:dencol}
\end{equation}
where the scaling parameter $A_{\Sigma} = 0.11$ and $b_{\Sigma} = 1/3$, which is also close to the value reported for the 3D density distribution in the isothermal case . 

On the other hand, Ballesteros-Paredes et al. (2011) numerically studied the evolution of the $\Sigma$--PDF resulting from colliding self-gravitating flows in presence of TI, which are intended to model the formation process of molecular clouds. They found that a very narrow lognormal regime appears when the cloud is being assembled. However, as the global gravitational contraction occurs, the initial density fluctuations are enhanced, resulting, first, in a wider lognormal $\Sigma$--PDF, and later, in a power-law $\Sigma$--PDF. These results suggest an explanation to the observational fact that clouds without star formation seem to possess a lognormal distribution, while clouds with active star formation develop a power-law tail at high column densities (e.g  Kainulainen et al. 2009).

In this work we quantitatively study  the behavior of the density and the column density distributions resulting in thermally bistable turbulent flows. For this purpose we analyze simple numerical 3-dimensional experiments which include only a cooling function appropriate for the diffuse neutral interstellar gas and turbulent forcing. The paper is organized as follows:  The model we use is described in section \S\ref{sec:model}. Then in section \S\ref{sec:results} we present the analysis of the behavior of both the volume density PDF and the column density PDF. These results are discussed in section  \S\ref{sec:disc}. Finally in  \S\ref{sec:conc} we present our conclusions.

\section{The model}\label{sec:model}
We use the same model as in Gazol \& Kim (2010), where a  MUSCL-type scheme (Monotone Upstream-centered Scheme for Conservation Laws) with HLL Riemann solvers (Harten, Lax, \& van Leer 1983; Toro 1999) is employed to solve hydrodynamic equations in three dimensions within a cubic computational domain with a physical scale of 100$\,$pc by side. The turbulence is randomly driven in Fourier space at large scales corresponding to $1 \leq k  \leq 2$, where $k$ is the magnitude of the wave vector, $k$. We use a purely solenoidal forcing because of two reasons. First, it is the most common kind of forcing used in the isothermal case and second, also for the isothermal case is the kind of forcing for which the density PDF is better described by a lognormal distribution (Federrath et al. 2008). 
For further details concerning the turbulent forcing see Gazol \& Kim (2010) and for a discussion on the effects of forcing see \S\ref{sec:for}. In the real ISM, turbulent motions are driven by spatially localized sources, as  the forcing in Fourier space is applied at any time in all the grid points, this kind of driving is not very realistic. However we chose this kind of driving method because it allows us to compare our results with previous work done for the isothermal case which includes Fourier space forcing. As an additional advantage, with this method we can study the density distribution for the dense gas as well as the density distribution for the diffuse gas (see \S\ref{sec:results}). We utilize the radiative cooling function presented by  S\'anchez-Salcedo et al. (2002), which is based on the standard $P$ vs. $\rho$ curve of Wolfire et al.\ (1995).

For all simulations we present in the next section the resolution is $512^3$, the boundary conditions are periodic, the gas is initially at the rest, and the initial density and temperature are uniform with $n_0=1\,$cm$^{-3}$ and $T_0=2399\,$K, which correspond to thermally unstable gas.  

\section{Results}\label{sec:results}
%
%
 In this section we analyze the density distribution resulting from one set of seven simulations with different  Mach number. The values of $M$ for those simulations are 0.28, 0.73, 1.86, 2.60, 3.29, 5.50, and 6.23, where $M$ is computed as the mean value of the local Mach number. In what follows we call this value $\langle M \rangle$ and  we use $M$ as a generic abbreviation for Mach number in situations that are independent of the specific way used to compute it or in situations where Mach numbers calculated in different ways are included. 

\subsection{The Volume Density Distribution}\label{sec:den_vol}
\begin{figure}
\plotone{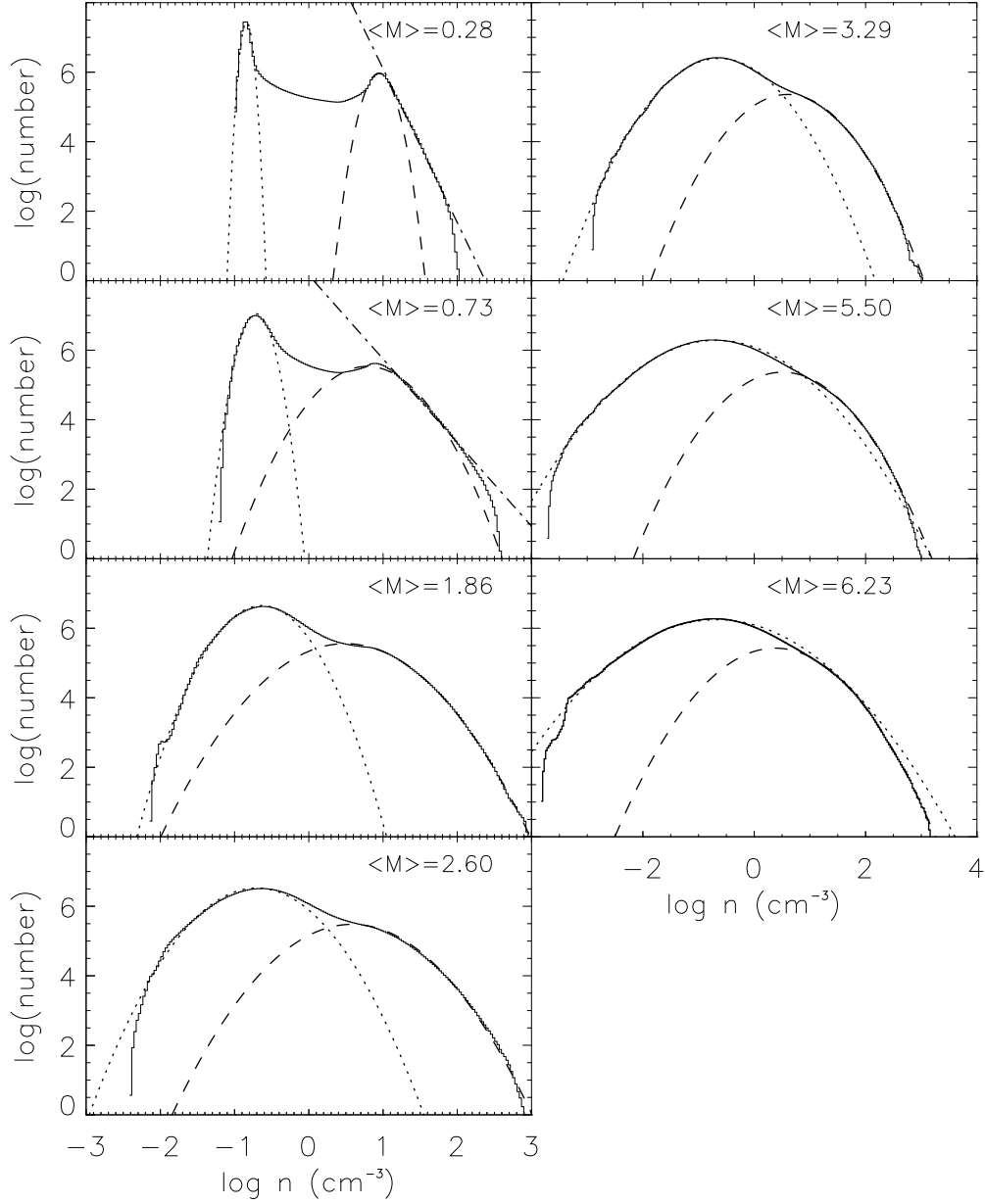}
\caption{Density PDF for different values of $\langle M \rangle$. {\it Dashed} and {\it dotted lines}  represent lognormal fits to the high and low density parts of the distribution, respectively; while {\it dashed-dotted lines}  in low $\langle M \rangle$ PDFs correspond to power-law fits with logarithmic slopes of $-4.43$ and $-2.41$ for $\langle M \rangle=0.28$ and $\langle M \rangle=0.73$, respectively.}
\label{fig:densidadN512}
\end{figure}

Volume density histograms resulting from these simulations are displayed in Figure \ref{fig:densidadN512}  along with lognormal fits to the high and the low density parts of the distribution. 
As expected from previous works (e.g. Gazol et al. 2005; Seifried et al. 2011), when the Mach number increases the PDF  becomes wider and  its bimodal nature, which is a consequence of TI development, becomes less pronounced. These two facts have been explained as a consequence of the decrease in the local ratio between the turbulent crossing time and the cooling time $\eta$ produced by the increase of $M$ (Gazol et al. 2005). This behavior have been proved by Seifried et al. (2011), who using test particles measured the time spent by a particle in the unstable regime as well as the frequency with which a test particle is perturbed, finding that both quantities increase with $M$ (i.e. both quantities increase as $\eta$ decreases). 
Concerning the shape of each part of the distribution, there are four things to note by a simple inspection. First, the low density zone of the PDF can be relatively well described by a lognormal (see dotted lines in Figure \ref{fig:densidadN512}). Second, the width of this lognormal increases with $\langle M \rangle$. Third, for the smallest values  of $\langle M \rangle$ we consider, the zone of the distribution corresponding to high densities is not well described by a lognormal (see dashed lines in Figure \ref{fig:densidadN512}). Fourth, when a lognormal  is an appropriate fit to the high density part, its width does not seem to systematically increase with the value of $M$ as rapidly as in the diffuse case. 

\begin{figure}
\plotone{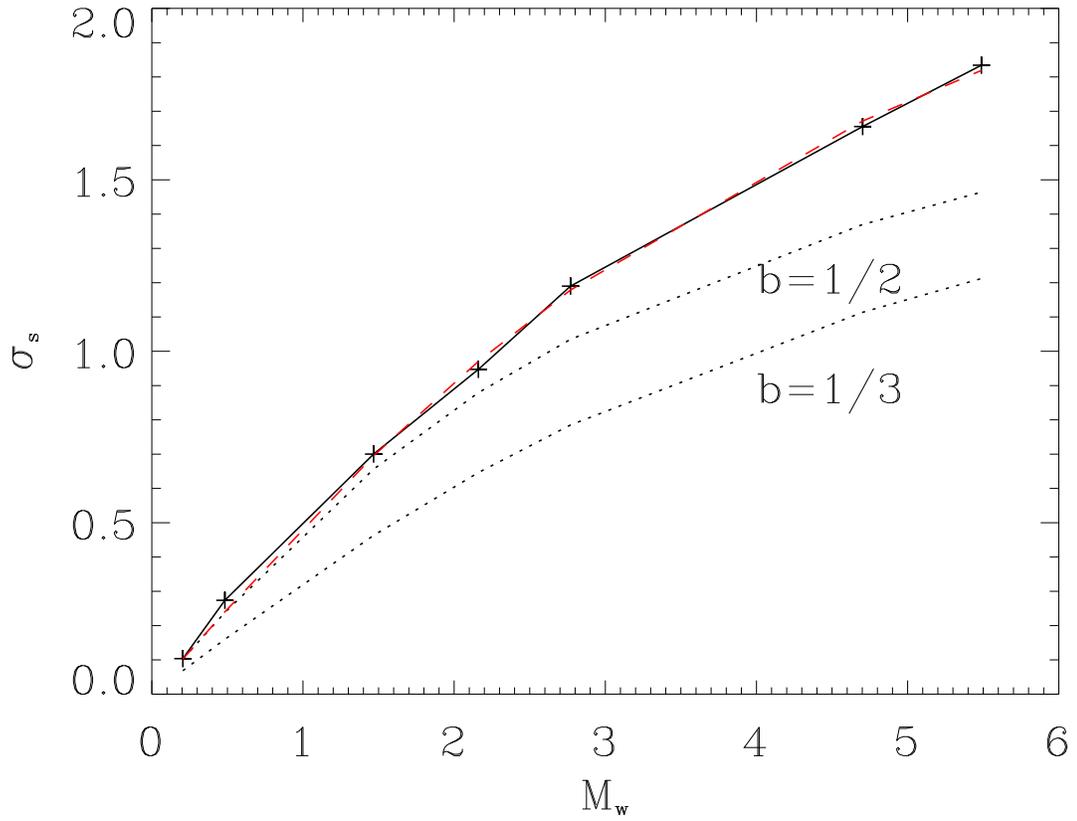}
\caption{Lognormal widths as a function of $M_{\rm w}$ for fits corresponding to low density gas in PDFs displayed in Figure \ref{fig:densidadN512}. {\it Dashed black lines}  plot $\sigma_s=(\ln (1+b^2M^2))^{1/2}$ for $b=1/2$ and $1/3$, whereas
{\it dashed red line} corresponds to the fit described in the text. Mach number values are average values of the local Mach number for the diffuse gas. }
\label{fig:anchosw}
\end{figure}

In Figure \ref{fig:anchosw} we show the width of the lognormal fits $\sigma_s$ (note that $\sigma_s$ is the width of the logarithmic density distribution $\ln n/\bar n$) that we obtain for the gas at low densities as a function of the local Mach $M_{\rm w}$, which is computed as the average value of the local Mach number in points with densities within the range where the fit has been computed. The dotted lines in this figure are plots of the relationship, between $\sigma_s$ and $M$, for isothermal gas (see eqn. \ref{sigmamach3d}) for two different values of the parameter $b$. Although for low $M_{\rm w}$ values, $\sigma_s$ is close to the $b=1/2$ isothermal case, for our simulations it grows faster with $M_{\rm w}$. In fact, {\it red} line represents a  fit with the form
\begin{equation}
\sigma_s^2=A\ln(1+b^2M^2),
\label{eqn:den}
\end{equation}
where $A=2.25$ and $b=0.33$. This value of $b$ is close
to typical values found for isothermal turbulence with solenoidal forcing.  

\begin{figure}
\plotone{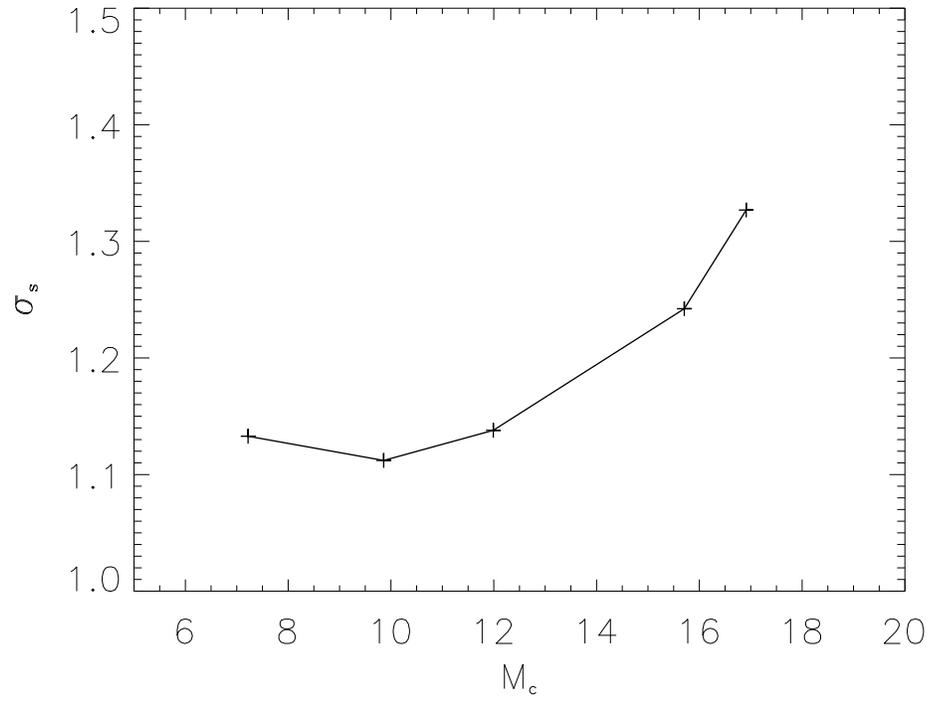}
\caption{Lognormal widths as a function of $M_{\rm c}$ for fits corresponding to the high density gas and $\langle M \rangle\geq 1.86$ in PDFs displayed in Figure \ref{fig:densidadN512}.}
\label{fig:anchosc}
\end{figure}

For the dense gas the behavior of $\sigma_s$ with $M_{\rm c}$ (computed as the average value of the local Mach number in points with $n>7.1\,$cm$^{-3}$, this density corresponds to the thermal equilibrium value below which the cooling function implies that the gas is thermally unstable), is qualitatively different from equation (\ref{eqn:den}) (see Fig. \ref{fig:anchosc}). In this case, the distribution is narrower than in the isothermal case and its width grows more slowly with $M_{\rm c}$ than in the isothermal case. In fact, due to the large difference between the widths we obtain and the ones implied by equation (\ref{eqn:den}) we do not include both cases in the figure. For instance, for  $M=10$ in the isothermal case $\sigma_s=1.58$ is expected if $b=1/3$.

\begin{figure}
\plotone{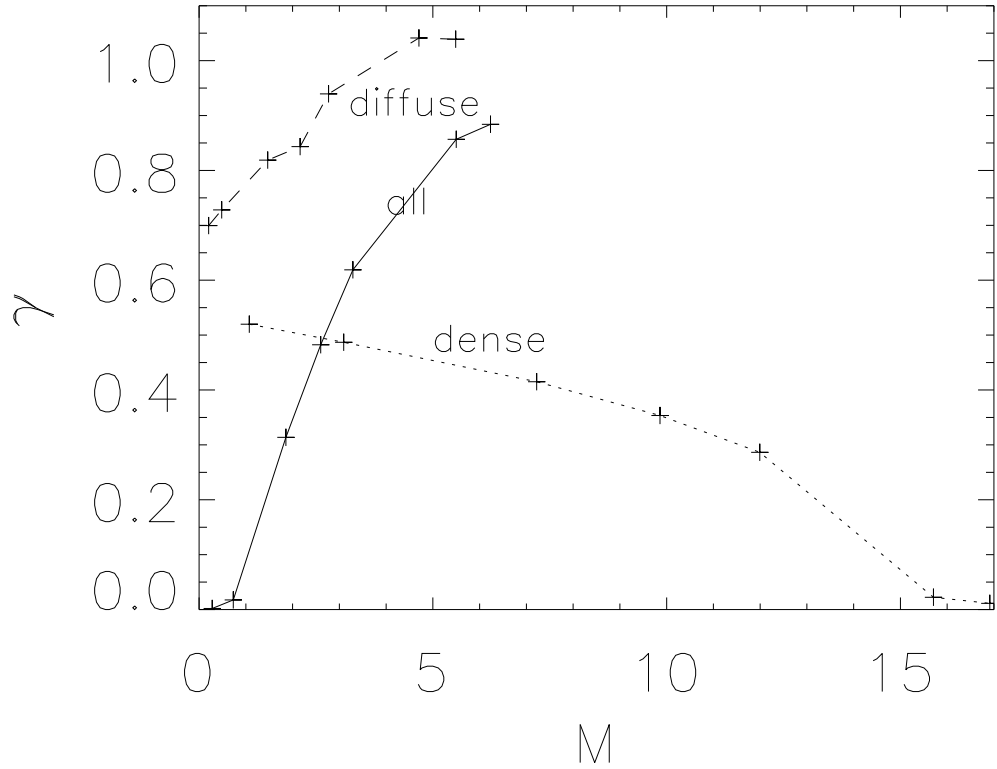}
\caption{Effective polytropic index $\gamma$ as a function of the corresponding averaged local Mach number for the whole simulation ({\it solid line}, $\langle M \rangle$), the diffuse gas ({\it dashed line}, $M_{\rm w}$) and the dense gas ({\it dotted line}, $M_{\rm c}$).}
\label{fig:pendientes}
\end{figure}

\begin{figure}
\plottwo{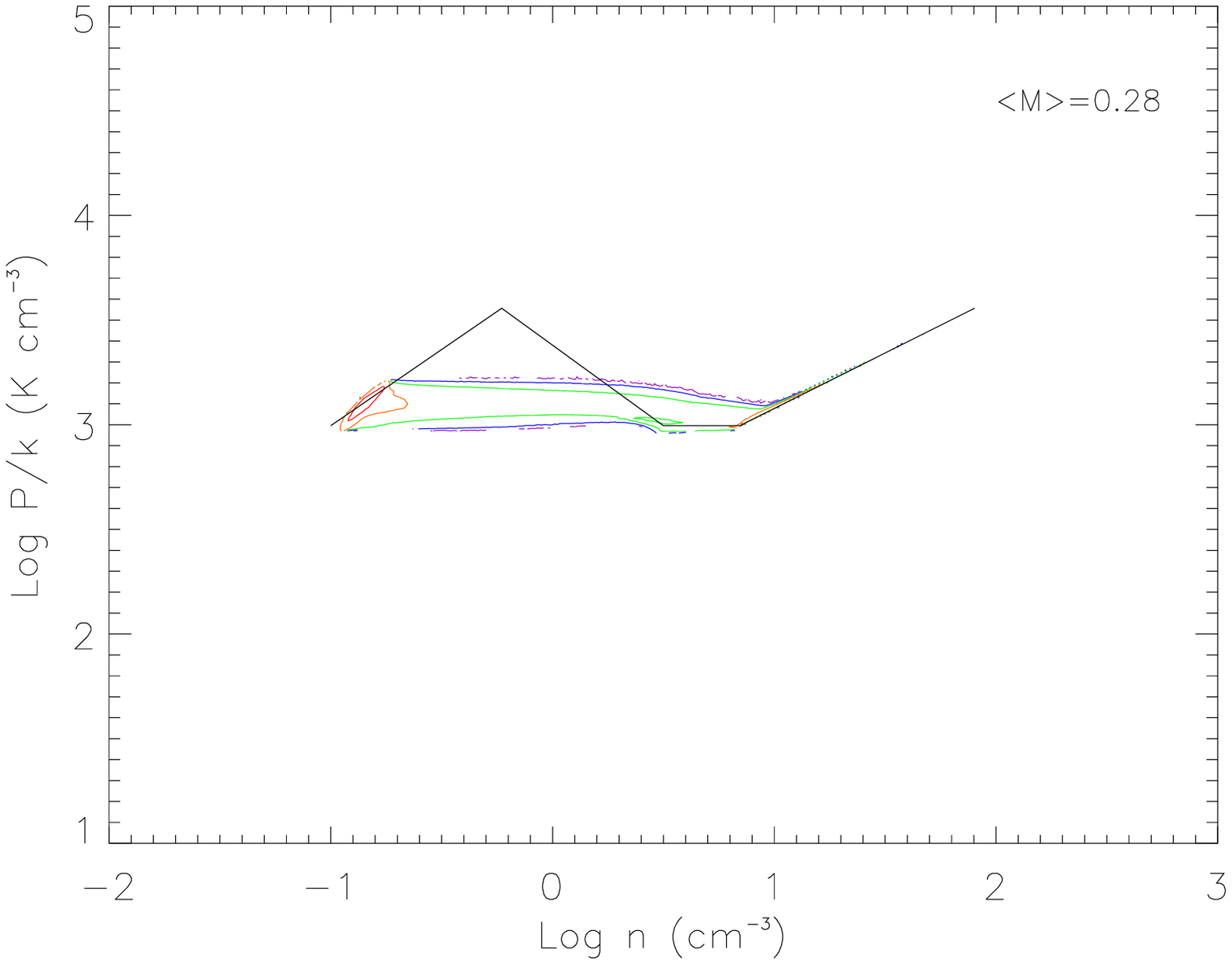}{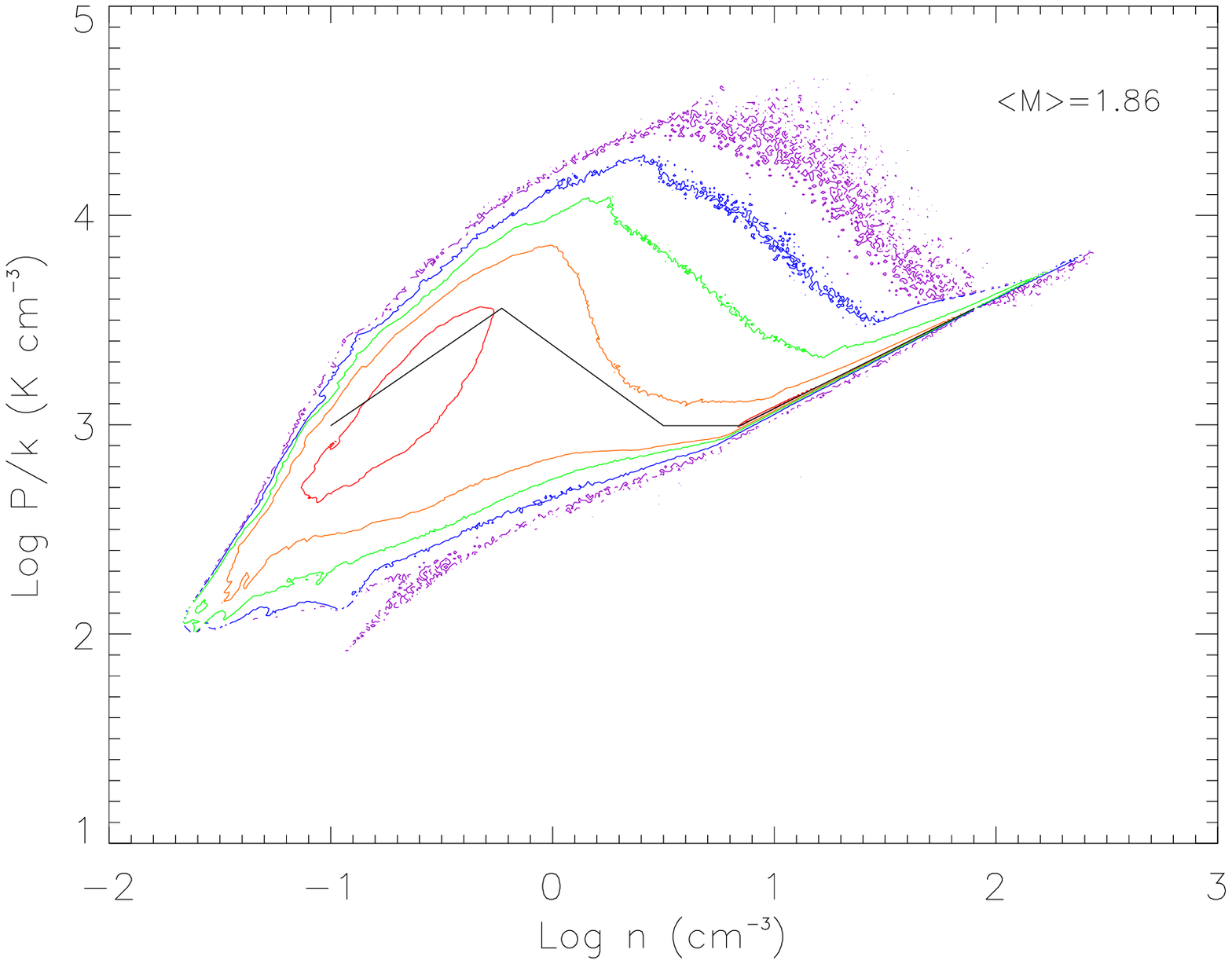}
\caption{Thermal pressure-density relation for simulations with $\langle M \rangle=0.28$ ({\it left}) and 1.86 ({\it right}). In 
each panel the black curve corresponds to the thermal equilibrium implied by the cooling function and contours are placed at 10$\%$ ({\it violet}), 30$\%$ ({\it blue}), 50$\%$ ({\it green}), 70$\%$ ({\it orange}), and 90$\%$ ({\it red}) of the logarithm of the maximum value of the two dimensional histogram.}
\label{fig:pvsrho}
\end{figure}

As mentioned in \S \ref{sec:intro}, previous works have found that the shape of the density PDF in non-isothermal turbulent flows depends on the value of the effective polytropic index $\gamma$ (Passot \& V\'azquez Semadeni, 1998). For our simulations we first measured this index as a line least squares slope of the whole $\log P$~vs.~$\log n$ distribution and we plotted it as a function of $\langle M \rangle$ (Fig. \ref{fig:pendientes}, {\it solid line}). 
As expected from Gazol et al. (2005), $\gamma$ increases with $\langle M \rangle$ and remains $< 1$.  
As can be seen from the $\log P/k$~vs.~$\log n$ distributions displayed in Figure \ref{fig:pvsrho}, for low values of $\langle M \rangle$ TI can develop almost without disturbance and the dense as well as the diffuse gas are predominately in thermal equilibrium with some of it transiting isobarically between stable branches. On the other hand, at high $\langle M \rangle$ the mean pressure in low density gas decreases below the thermal equilibrium value while the mean pressure in high density gas increases above its thermal equilibrium value, producing a neat positive slope of the whole $\log P/k$~vs.~$\log n$ distribution. 
If the value of $\gamma$ obtained in this way were a suitable parameter to infer the shape of the density PDF, we would expect a power law at high density for all our simulations. Also from Figure \ref{fig:pvsrho} it is clear that for low values of $\langle M \rangle$ almost all the dense gas ($n > 7.1\,$cm$^{-3}$) is in thermal equilibrium,  implying that its thermodynamic state is approximately well described by a polytropic relation with $\gamma = 0.53$, which is the power corresponding to the thermal equilibrium curve and which  accordingly with the Passot \& V\'azquez-Semadeni (1998) theory should exhibit a power-law PDF. The  {\it dotted line} in Figure \ref{fig:pendientes} corresponds to the slope of the $\log P/k$~vs.~$\log n$ distribution computed for densities larger than $7.1\,$cm$^{-3}$ and plotted as a function of the average local Mach number for dense gas $M_{\rm c}$.  This slope shows a decrease with $M_{\rm c}$ from $\sim 0.53$  for $\langle M \rangle=0.28$ ($M{\rm c}=1.07$) to  $\sim 0$ for $\langle M \rangle=6.23$ ($M{\rm c}=16.91$). This decrease is due to the increased dispersion at the lower density values of high density gas, and  implies that the polytropic description of the dense gas becomes less appropriate for large $M$ values.
For diffuse gas the slope of the distribution ({\it dashed line}), estimated for each simulation using density values within the range for which the fit to the corresponding PDF has been computed, increases from its thermal equilibrium value, namely $0.73$, for low $M_{\rm w}$ simulations to approximately $1$ for very supersonic simulations, implying that the approximately lognormal shape we obtain is consistent with the predictions done by Passot \& V\'azquez-Semadeni (1998).   

\subsection{The Column Density Distribution}\label{sec:den_col}

 As long as the authors know, the column density distribution resulting from turbulent thermally unstable simulations, has not been systematically studied. It is thus interesting to study the $\Sigma$--PDF associated with the $n$--PDFs discussed in previous section.

\begin{figure}
\plotone{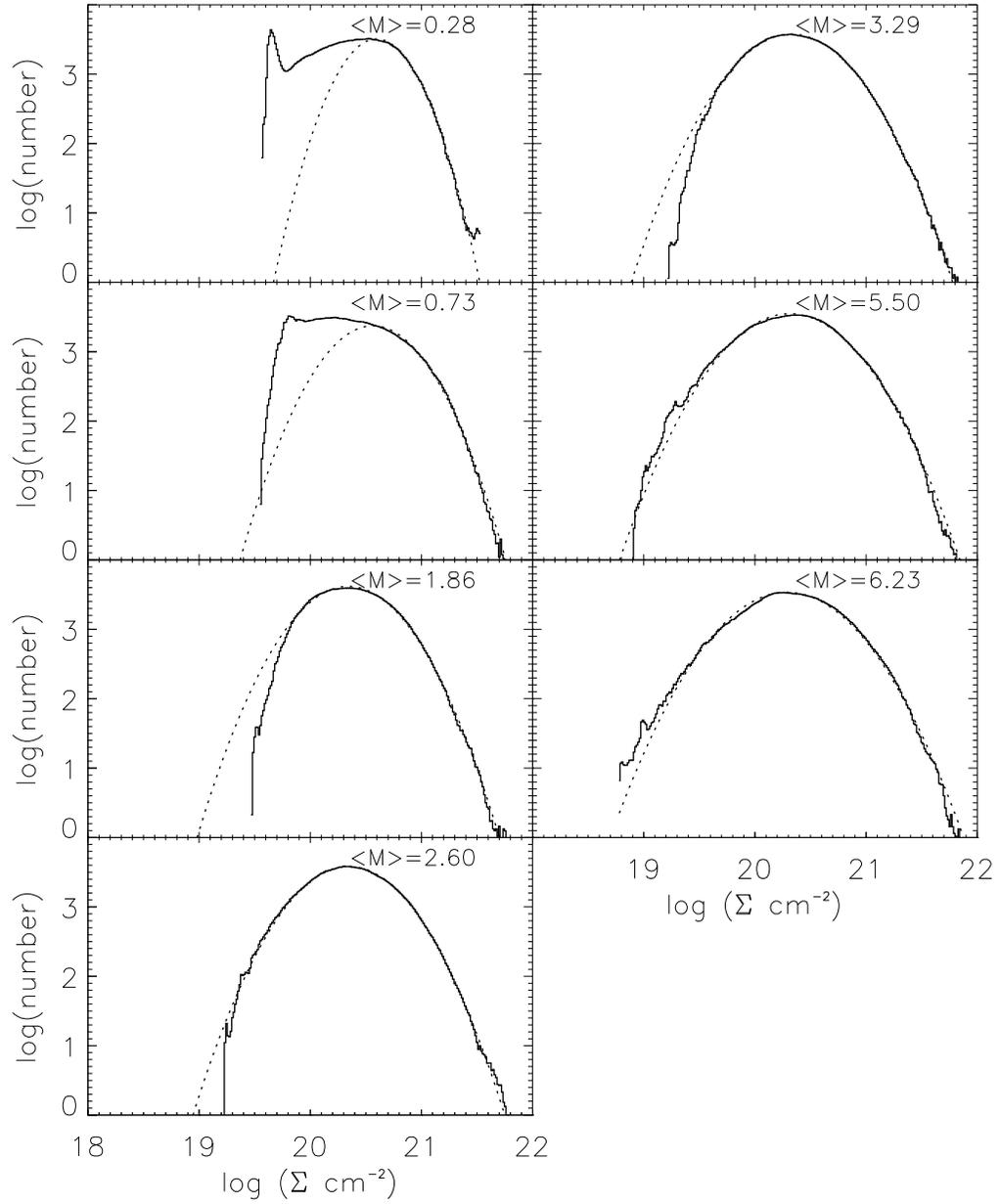}
\caption{Column density PDF for different values of $\langle M \rangle$. {\it Dotted lines} represent lognormal fits.}
\label{fig:densidadcolN512}
\end{figure}

\begin{figure}
\plotone{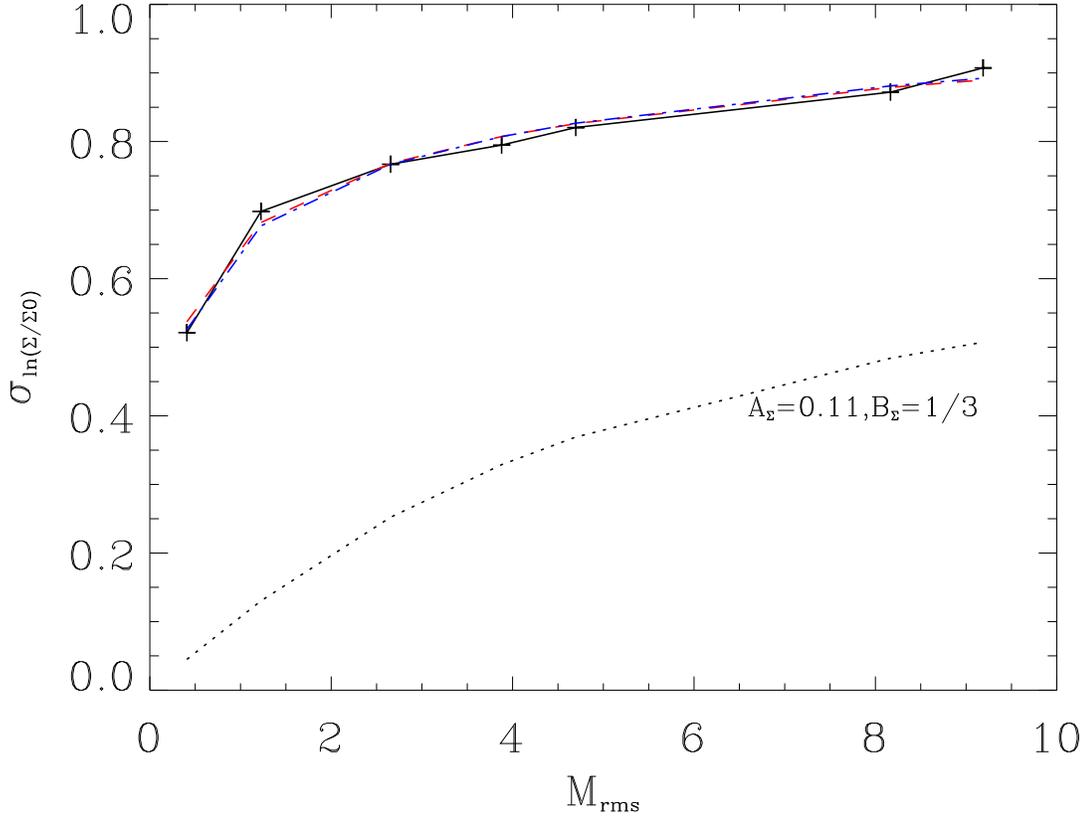}
\caption{Lognormal widths as a function of $M_{\rm rms}$ for fits corresponding to column density distributions displayed in Figure \ref{fig:densidadcolN512}. In this plot Mach number values are rms values at the mean temperature. {\it Dashed black lines}  plot equation (\ref{eqn:dencol}) for $A_{\Sigma} = 0.11$ and $b_{\Sigma} = 1/3$, whereas
{\it dashed red} and {\it dashed blue lines}  correspond to the fits described in the text. Mach number values are the rms values at the mean temperature $M_{rms}$. }
\label{fig:anchoscol}
\end{figure}

In Figure \ref{fig:densidadcolN512} we show the column density distribution resulting from our set of simulations. The high density part of the distribution can be well described by a lognormal ({\it dotted lines}) for all the values of $\langle M \rangle$ that we consider, even for the two smaller values (0.28 and 0.73) for which de volume density PDF has a behavior approaching a power-law. For these two values the bimodal nature of the $n$--PDF is preserved in the  $\Sigma$--PDF. For larger values of $\langle M \rangle$  the distribution becomes single peaked and the lognormal does also fit an important fraction of the low density part. The widths of the lognormals,  plotted as a function of the Mach number in Figure \ref{fig:anchoscol} ({\it solid line}), are systematically larger than the corresponding values found by Burkhart \& Lazarian (2012) for the isothermal case ({\it dotted line}), but they follow a qualitatively similar behavior with $M$. In fact this behavior can be fit by a function of the form (\ref{eqn:dencol}) with $A_{\Sigma}=0.084$ and $b_{\Sigma}=12.5$ ({\it dashed red line}) obtaining an error in the fit of 2.17$\%$. This set of parameters is however not unique. As an example we also display the curve for $A_{\Sigma}=0.081$ and $b_{\Sigma}=14.29$ ({\it dashed blue line}) which has an error of 2.22$\%$. Considering the fact that errors in fiting lognormals are greater than the difference between the errors resulting from  the two previous fits, we can consider them as being equivalent. Note that in Figure \ref{fig:densidadcolN512} the Mach number is the rms value at the mean temperature $M_{rms}$. We choose this value because the average local Mach is dominated by the warm gas ($T>6100$K), whose volume fraction is between $\sim 30\%$, for very turbulent simulations, and $\sim 80\%$, for the smaller value of $M$,  which is very large when compared with the one of the cold gas ($T<310\,$K) which goes from $\sim 2$ to $8\%$. 

\section{Discussion}\label{sec:disc} 
\subsection{Diffuse Gas}\label{sec:discdiff}
The approximate lognormal shape of the diffuse gas density PDF is consistent with observations reported by Berkhuijsen \& Fletcher (2008) for galactic low density neutral gas. However, the fact that the lognormal fit fails at very low densities is also consistent with predictions by Passot \& V\'azquez-Semadeni (1998) for polytropic flows with $\gamma < 1$,  according to which the PDF is expected to decrease faster than a lognormal at low densities.

 The behavior of $\sigma_s$ with $M$, that grows faster than in the isothermal case, shows the effects of the presence of TI. In fact as discussed in V\'azquez-Semadeni et al. (2003), the increase of $M$ has two effects on the development of TI. First, as the turbulent crossing time decreases it becomes shorter than the growth time for linear TI, which is the time for the gas to form condensations, increasing the fraction of unstable gas (e.g. Gazol et al 2001) and producing a larger drift from the thermal equilibrium. On the other hand, the presence of velocity fluctuations generate adiabatic density perturbations which are linearly unstable only when the cooling time is shorter than the dynamical time (which for the supersonic case is the turbulent crossing time), i.e. only at large scales . However, in the nonlinear regime, reached as $M$ increases,  due to the local increase of density in forcing generated compressions, the cooling time can locally decrease implying that even initially small scale fluctuations can become thermally unstable. The consequence of this  nonlinear development of TI is an enhancement of density contrast that could lead the width of the density PDF to grow faster with $M$ than in the isothermal case.

\subsection{The dense Gas}\label{sec:discdense}

The results we obtained for the high density gas reconcile the theoretical prediction from Passot \& V\'azquez-Semadeni (1998)  with previous numerical results for thermally bistable flows (Audit \& Hennebelle 2010). Following the former, for a cooling function adapted to describe the dense atomic gas, having an effective polytropic index $\gamma<1$ in thermal equilibrium, the density PDF is expected to develop a power-law at high values. On the other hand Audit \& Hennebelle (2010) report that the dense gas density PDF resulting from thermally bistable colliding flows seems to be better described by a lognormal than by a power-law. From results presented in \S\ref{sec:den_vol}  it is clear that the dense gas PDF can behave as a power-law or as a lognormal depending on the $M$ value and more specifically on the amount of dense gas out of thermal equilibrium. The transition between these behaviors could be due to the fact that the higher is the amount of gas out of thermal equilibrium, the lesser is adequate a polytropic equation of state as a thermodynamic description of the gas. In fact, Passot \& V\'azquez-Semadeni (1998) suggest that the development of a power-law for $\gamma\neq 1$ is the consequence of density jumps depending the local density. In our simulations the dense gas drifting away from thermal equilibrium is in fact the consequence of the density fluctuations being determined by velocity fluctuations. A remarkable difference between the power-laws we obtain at high densities for low $M$ simulations and the predictions from Passot \& V\'azquez-Semadeni (1998) is the logarithmic slope behavior. We get slopes of  $-4.43$ and $-2.41$ for $\langle M \rangle=0.28$ and $\langle M \rangle=0.73$, respectively. This implies that the power-laws resulting from our simulations are stepper than the power-law they predict for $\gamma=0.5$ which is expected to have a logarithmic slope of $-1.2$. Note however that this value corresponds to the large $M$ limit. 

The development of a lognormal at high densities in high $M$ simulations suggests that in some cases the use of this kind of distribution as initial condition in the molecular cloud formation process represents an adequate choice. Nevertheless,  for our simulations the relationship  $\sigma_s$-$M$ is very different from equation (\ref{eqn:den}), implying that in early times during the cloud formation process the density distribution can not be related with the dynamical state of the gas through the isothermal version of this equation. In particular, for a given Mach number we find a narrower distribution. 
It is important to note that the presence of self gravity, which is not included in our models, could  have noticeable effects on the distribution of dense gas even in these early phase. This problem is going to be addressed in a future work.

\subsection{The column density}\label{sec:disccol}   

The dense branch of the column density PDF is well described by a lognormal regardless of the $M$ value. This is partially consistent with recent observational and numerical works suggesting that in clouds where the effects of gravity are not dominant in determining the cloud structure, the column density PDF has a lognormal shape (Kainulainen et al. 2011, Ballesteros-Paredes et al. 2011). In those works however, turbulence is invoqued as the main agent in shaping the column density PDF. From our results it is possible to suggest that even when low levels of turbulence are present, the cooling properties of the gas could produce a lognormal column density PDF. 
In fact, Heitsch et al. (2008) investigate the expected timescales of the dynamical and thermal instabilities leading to the rapid fragmentation of gas swept up by large-scale flows and compare them with global gravitational collapse timescales. They identify parameter regimes in gas density, temperature, and spatial scale within which a given instability dominate, finding that  the thermally dominated parameter regime has a remarkable large extent due to the fact that outside the strictly thermally unstable regions, cooling could still being the dominating agent  leading to fragmentation  in the presence of  an external (in this case ram) pressure.

On the other hand we find that the width of the column density distribution is much larger than in the isothermal case studied by other authors for MHD flows (Kowal, Lazarian \& Bresnyak 2007; Burkhart \& Lazarian 2012). This is a natural consequence of the large density dynamical range produced by the presence of thermal instability. The resulting  $\sigma_{ln(\Sigma/Sigma_0)}$-$M$ relationship has to include the fact that the density contrast is large even for low Mach numbers requiring  a very rapid grow that implies parameter values completely different from the ones reported in the isothermal case, namely  $A_{\Sigma}=0.11$ and $b_{\Sigma}=1/3$.   

\subsection{Numerical Issues}\label{sec:num}

Several numerical factors can affect in some way the results obtained in the present work. 
\subsubsection{Resolution}\label{sec:res}  
We have performed some higher resolution simulations in order to see the effects on the density PDF. For similar Mach numbers we find that the main consequence of increasing the resolution to $1024^3$ seems a deviation of the high density tail in moderate Mach simulations. In particular, for $\langle M \rangle\sim 1.9$, the distribution falls faster than a lognormal for high $n$. Unfortunately we do not have enough snapshots to quantify this effect and to measure $\sigma$ in high resolution simulations. Although our PDFs could be not fully converged and higher resolution simulations could potentially lead to results quantitatively different from those presented in previous sections, the facts that the diffuse gas PDF can be well described by a lognormal and that the dense gas PDF is well described by this kind of distribution only for large enough values of $M$, does not seem to change with resolution. 
\subsubsection{Model}\label{sec:mo}
Even if our model allows both, the study of diffuse gas distribution and a direct comparison with isothermal numerical experiments reported by other authors, there are some choices which can potentially affect our results. The periodic boundary conditions maintain a fixed amount of gas in the box and this fact can artificially regulate the gas segregation. Also the  particular choice of the cooling function can affect the gas segregation. This function could change because of physical reasons such as abundance variations, heating rate variations, or additional cooling process.  Finally, additional physics such as the presence of self gravity and magnetic fields could also modify the quantitative behavior of the density distribution. 
\subsubsection{Forcing}\label{sec:for} 
Federrath et al. (2008) and Federrath et al. (2010) found  that for isothermal gas the width of the PDF depends not only on the rms Mach number but also on the relative degree of compressible and solenoidal modes in the turbulence forcing, with $b = 1/3$ appropriate for purely solenoidal and $b = 1$ for purely compressive forcing. For the thermally bistable case the effects of forcing could be more complex due to the  interplay between TI and turbulence. Although simulations with different kind of forcing have been presented by Seifried at al. (2011), the energy transfer between the solenoidal and the compressive modes for turbulent bistable flows have not been addressed.  As shown in Gazol \& Kim (2010) for purely solenoidal forcing, the presence of TI does significantly affect the density as well as the velocity power spectrum. In fact, it is well known that the development of TI produce turbulent motions with typical velocities of the order of tenths of km s$^{-1}$ (Kritsuk \& Norman 2002; Piontek \& Ostriker 2004) but, as stated earlier, the presence of turbulence does in turn modify the development of TI. A detailed analysis of the effects of forcing comparing also the resulting velocity power spectrum and the density weighted velocity power spectrum for the compressible as well as for the solenoidal modes, should be done in order to address this problem, but it is out of the scope of the present work.  

\section{Summary and Conclusions}\label{sec:conc}

In the present work we have presented numerical experiments showing that:
\begin{enumerate}
\item At low densities the volume density PDF resulting from turbulent thermally bistable flows can be well described by a lognormal distribution whose width increases with the Mach number.     
\item A relationship between the width of the distribution and the average local Mach number can  be found, however this relationship is not the same as in the isothermal case. The value of the parameter $b$, included in the isothermal case, is for our simulations surprisingly close to the one obtained for purely solenoidal forcing in isothermal gas but the non linear development of TI, producing a more efficient rarefaction of gas, causes a faster growth of the distribution width as $M$ is increased. The consequence of this enhanced growth in the mathematical form of the $\sigma_s-M$ relation is the presence of a scale factor distinct from 1.
\item At high densities the form of the volume density PDF depends on the value of the Mach number. For simulations with transonic or weakly supersonic average velocities in dense gas the distribution is a power-law, while in the presence of highly supersonic velocities the distribution becomes lognormal.
\item For simulations which develop a high density distribution with a lognormal shape, the witdth of the distribution is smaller than in the isothermal case and grows slower with the Mach number. 
\item At high densities the column density PDF resulting from our simulations can be described by a lognormal for all the Mach numbers we consider. As $M$ increases the density range where the lognormal fit is adapted expands and the lognormal becomes wider.
\item The width of the column density distribution resulting from our simulations is systematically larger than the width obtained in the isothermal case. A relationship between the width of the column density distribution and the rms Mach number at the mean temperature can be found. This relationship has the same form as the one reported in the literature for the isothermal case, but the parameters resulting from our fit are very different. 

\end{enumerate} 

From these results it is clear that when studying the diffuse and/or the dense atomic interstellar gas in order to relate the density structure, and in particular the width of its PDFs, with the dynamical state of the gas, characterized by the Mach number, the use of results obtained from isothermal turbulent flows is not an adequate choice. Specific relations between $\sigma_s$ ($\sigma_{ln(\Sigma/\Sigma_0)}$) and $M$ for thermally bistable flows should be taken into account in any observational or theoretical work using the density (column density) PDF as a measure of the Mach number. The relationships obtained in the present work could be affected by the inclusion of additional physics such as self gravity, magnetic fields or variations on the cooling function due to variations of the heating rate and the gas abundances.

\acknowledgements 
The work of A. G. was partially supported by UNAM-DGAPA grant IN106511
 Some of the numerical simulations were performed 
at the cluster Platform 4000 (KanBalam) at DGSCA, UNAM.

\bibliographystyle{apj}

\begin{thebibliography}{}

\bibitem[Audit \& Hennebelle (2010)]{2010A&A...511A..76A}
Audit, E., \& Hennebelle P. 2010 \aap, 115, 76

\bibitem[Ballesteros-Paredes et al. (2011)]{2011MNRAS..}
Ballesteros-Paredes, J., V\'azquez-Semadeni, E., Gazol A., Hartmann, L., Heischt, F., \& Col{\'\i}n P. \ 2011, \mnras, 1436, 416

\bibitem[Beetz et al.(2008)]{2008PhLA..372.3037B} 
Beetz, C., Schwarz, C., Dreher, J., \& Grauer, R.\ 2008, Physics Letters A, 372, 3037 

\bibitem[ Berkhuijsen \& Fletcher (2008)]{2008MNRAS.390L..19B} Berkhuijsen, E. M., \& Fletcher, A. 2008, \mnras, 390L, 19

\bibitem[Boldyrev et al. (2002)]{2002PhRvL..89c1102B} 
Boldyrev, S., Nordlund, {\AA}., \& Padoan, P.\ 2002, Physical Review Letters, 89, 031102 

\bibitem[Burkhart \& Lazarian (2012)]{BurLaz12}
Burkhart, B., \& Lazarian, A. 2012, \apj, 755,  L19

\bibitem[deAvillez \& Breitschwert (2005)]{2005A&A...436A..585A}
de Avillez, M. A., \& Breitschwerdt, D. 2005, \aap, 436, 585

\bibitem[Diplas \&  Savage (1994)]{1994ApJS...93..211D}Diplas, A., \&  Savage, B. D.\  1994, \apjs, 93, 211

\bibitem[Elmegreen (2002)]{2002ApJ...577..206E} Elmegreen, B.~G.\ 2002, 
\apj, 577, 206 

\bibitem[Elmegreen (2008)]{2008ApJ...672.1006E} Elmegreen, B.~G.\ 2008, 
\apj, 672, 1006 

\bibitem[Elmegreen (2011)]{2011ApJ...731...61E} Elmegreen, B.~G.\ 2011, 
\apj, 731, 61 

\bibitem[Federrath et al. (2008)]{2008ApJ...688L..79F} Federrath, C., 
Klessen, R.~S., \& Schmidt, W.\ 2008, \apjl, 688, L79 

\bibitem[Federrath et 
al.(2010)]{2010A&A...512A..81F} Federrath, C., Roman-Duval, J., Klessen, R.~S., Schmidt, W., \& Mac Low, M.-M.\ 2010, \aap, 512, A81 

\bibitem[Field (1965)]{Fie65}
Field, G. B. 1965,  \apj, 142, 531

\bibitem[Gazol \& Kim (2010)]{GK10}
Gazol, A., \& Kim,  J. 2010 \apj, 723, 482

\bibitem[Gazol et al. (2005)]{GVK05}
Gazol, A., \VS, E., \& Kim,  J. 2005 \apj, 630, 911

\bibitem[Harten et al. (1983)]{hll83}
Harten, A., Lax, P. D., \& van Leer, B. 1983, SIAM, Rev., 25, 35

\bibitem[Heitsch et al. (2008)]{hhb08}
Heitsch, F.,  Hartmann, L., \& Burkert A. 2008, \apj, 683, 786

\bibitem[Hennebelle \& Audit (2007)]{henau07}
Hennebelle, P., \& Audit, E. 2007,   \aap, 465, 431

\bibitem[Hennebelle \& Chabrier (2008)]{2008ApJ...684..395H} Hennebelle, P., \& Chabrier, G.\ 2008, \apj, 684, 395 

\bibitem[Hennebelle \& P\'erault(1999)]{hen99}
Hennebelle, P., \& P\'erault, M. 1999, \aap, 351, 309

\bibitem[Joung and Mac Low (2006)] {joungmaclow06}
Joung, M. K. R., \&  Mac Low, M. M.  2006 \apj, 653, 1266

\bibitem[Kainulainen et al. (2011)] {kainu11}
Kainulainen J., Beuther H., Banerjee R., Federrath C., \& Henning T.
2011, \aap, 530, A64

\bibitem[Kainulainen et al. (2009)] {kainu09}
 Kainulainen J., Beuther H., Henning T., \& Plume R. 2009, \aap, 508, L35

\bibitem[Kim et al. (1999)]{kim_etal99}
Kim, J., Ryu, D., Jones, T. W., \&  Hong, S. S. 1999, \apj, 514, 506

\bibitem[Klessen (2000)]{2000ApJ...535..869K} Klessen, R.~S.\ 2000, \apj, 
535, 869 

\bibitem[Kritsuk \& Norman (2002)]{kn02}
Kritsuk, A., \& Norman, M. L. 2002, \apj, 569, L127

\bibitem[Krumholz \& McKee (2005)]{2005ApJ...630..250K}
 Krumholz, M.~R., \& McKee, C.~F.\ 2005, \apj, 630, 250 

\bibitem[Kowal, Lazarian \& Bresnyak (2007)]{klb07}
Kowal, G., Lazarian, A.,\&  Beresnyak, A. 2007,\apj , 658, 423

\bibitem[Lemaster \& Stone (2008)]{2008ApJ...682L..97L} 
Lemaster, M.~N., \& Stone, J.~M.\ 2008, \apjl, 682, 97 

\bibitem[Li et al. (2003)]{2003ApJ...592..975L}
Li, Y., Klessen R. ~S.,\&  Mac Low, M. ~M. 2003, \apj, 592, 975

\bibitem[Ostriker et al.(2001)]{2001ApJ...546..980O} Ostriker, E.~C., 
Stone, J.~M., \& Gammie, C.~F.\ 2001, \apj, 546, 980 

\bibitem[Padoan \& Nordlund (2002)]{2002ApJ...576..870P} 
Padoan, P., \& Nordlund, {\AA}.\ 2002, \apj, 576, 870 

\bibitem[Padoan \& Nordlund (2011)]{2011ApJ...730...40P} 
Padoan, P., \& Nordlund, {\AA}.\ 2011, \apj, 730, 40 

\bibitem[Padoan, Nordlund, Jones  (1997)]{pnj97} 
Padoan, P., Nordlund, \& Jones B. J. T.\ 1997, {\AA}, \mnras, 288, 145 

\bibitem[Passot \& V{\'a}zquez-Semadeni (1998)]{1998PhRvE..58.4501P} 
Passot, T., \& V{\'a}zquez-Semadeni, E.\ 1998, \pre, 58, 4501 

\bibitem[Popntek \& Ostriker (2004)]{PIOS04}
Piontek, R. A., \& Ostriker, E. C. 2004, \apj, 601, 905 

\bibitem[Price et al. (2011)]{2011ApJ...727L..21P} 
Price, D.~J., Federrath, C., \& Brunt, C.~M.,\ 2011, \apjl, 727, L21 

\bibitem[Robertson \&  Kravtsov (2008)]{robkrav08}
Brant E. Robertson, B. E., \&  Kravtsov, A. V. 2008, \apj, 680, 1083

\bibitem[Seifried & al. (2011)]{2011A&A...526A..14S}
Seifried, D. and Schmidt, W., \& Niemeyer, J.~C. 2011, \aap, 506, A14

\bibitem[Stone et al.(1998)]{1998ApJ...508L..99S} 
Stone, J.~M., Ostriker, E.~C., \& Gammie, C.~F.\ 1998, \apjl, 508, L99 

\bibitem[Tamburro et al.  (2009)]{tamb09}
Tamburro, D., Rix, H. W., Leroy, A, K., MacLow, M. M., Walter, F.,
Kennicutt, R. C., Brinks, E., \& de Blok, W. J. G. 2009 \aj, 137, 4424

\bibitem[Toro et al. (1999)]{toro99}
Toro, E. F. 1999, Riemann Solvers and Numerical Methods for Fluid Dynamics: A
practical Introduction (Berlin: Springer)

\bibitem[V\'azquez-Semadeni (1994)]{1994ApJ...423..681V} 
V\'azquez-Semadeni, E.\ 1994, \apj, 423, 681 

\bibitem[V\'azquez-Semadeni \& Garc{\'\i}a (2001)] {2001ApJ...557..727V}
V\'azquez-Semadeni, E., \& Garc{\'\i}a N. \ 2001, \apj, 557, 727

\bibitem[V\'azquez-Semadeni et al. (2003)]{vgps03} 
V\'azquez-Semadeni, E., Gazol, A., S\'anchez-Salcedo, F. J., and
Passot, T. ed. Thierry Passot \& Edith Falgarone, 2003, Lecture Notes in Physics, vol. 614, p.213-251

\bibitem[V{\'a}zquez-Semadeni et al. (2006)]{2006ApJ...643..245V} 
V{\'a}zquez-Semadeni, E., Ryu, D., Passot, T., Gonz{\'a}lez, R.~F., \& Gazol, A.\ 2006, \apj, 643, 245 

\bibitem[Wada & Norman (2007)]{wadanorman07}
Wada, K., \& Norman, C. A. 2007, \apj, 660, 276

\bibitem[Wolfire et al (2003)]{wol03}
Wolfire, M. G., McKee, C. F.,  Hollenbach, D., \& Tielens, A. G. G. M.
2003, 587, 278

\bibitem[Wolfire, Hollenbach \& McKee (1995)]{wol95}
Wolfire, M. G., Hollenbach, D., McKee, C. F., Tielens, A. G. G. M., \&
Bakes, E. L. O. 1995, \apj, 443, 152

\bibitem[Zamora-Avil\'es, V\'azquez-Semadeni, \&  Col{\'\i}n (2012)]{zam2012}
Zamora-Avil\'es, M., V\'azquez-Semadeni, E., \&  Col{\'\i}n, P. 2012, \apj, 751, 77

\end{thebibliography}

\epsscale{.85}

\epsscale{1.0}

\epsscale{.850}

\epsscale{1.0}

\epsscale{1.0}

\epsscale{.850}

\epsscale{1.0}

\end{document}